\documentclass[twoside]{article}
\usepackage[latin1]{inputenc}
\usepackage[T1]{fontenc}
\usepackage{parskip}
\usepackage{amsmath,amsfonts,amssymb,amsxtra}
\usepackage{hyperref}

\usepackage{latexsym}
\usepackage{theorem}
\usepackage{graphicx,psfrag}
\usepackage{verbatim}

{\theoremstyle{plain}                       
\theorembodyfont{\itshape}
\newtheorem{Theorem}{Theorem}}

{\theoremstyle{plain}                       
\theorembodyfont{\itshape}
\newtheorem{Corollary}{Corollary}}

{\theoremstyle{plain}                       
\theorembodyfont{\itshape}
\newtheorem{Proposition}{Proposition}}

{\theoremstyle{plain}                       
\theorembodyfont{\rmfamily}
\newtheorem{Definition}{Definition}}

{\theoremstyle{plain}                       
\theorembodyfont{\rmfamily}
}

{\theoremstyle{plain}                       
\theorembodyfont{\rmfamily}
\newtheorem{Example}{Example}}

{\theoremstyle{plain}                       
\theorembodyfont{\rmfamily}
\newtheorem{Remark}{Remark}}

{\theoremstyle{plain}                       
\theorembodyfont{\itshape}
\newtheorem{Lemma}{Lemma}}

\begin{document}

\title{ Dynamically defined measures and equilibrium states}
\author{Ivan Werner\\
   {\small Email: ivan\_werner@mail.ru}}
\date{December 15, 2011}
\maketitle

\begin{abstract}\noindent
A technique of dynamically defined measures is developed and its
relation to the theory of equilibrium states is shown. The technique
uses Carath\'{e}odory's method and the outer measure introduced in
\cite{Wer3}. As an application, equilibrium states for contractive
Markov systems \cite{Wer1} are obtained.


\end{abstract}

\section{Dynamically defined measures}

Let $E$ be a finite set and $\Sigma:=\{(...,\sigma_{-1},\sigma_0,\sigma_1,...):\sigma_i\in E\
\forall i\in\mathbb{Z}\}$ equipped with the product topology of the discrete topologies.  Let $S:\Sigma\longrightarrow\Sigma$ be the left shift map on $\Sigma$, i.e. $(S\sigma)_i = \sigma_{i+1}$ for all $i\in\mathbb{Z}$. We call the set $
_m[e_m,...,e_n]:=\{\sigma\in\Sigma:\ \sigma_i=e_i\mbox{ for all
}m\leq i\leq n\}$ for $m\leq n\in\mathbb{Z}$, a {\it cylinder}.
Let $\mathcal{A}$ denote the $\sigma$-algebra generated by the
zero time partition $\{_0[e]:e\in E\}$ of $\Sigma$, and define, for
each integer $m\leq 1$,
\[\mathcal{A}_m:=\bigvee\limits_{i=m}^{+\infty} S^{-i}\mathcal{A},\]
which is the smallest $\sigma$-algebra containing  all
 $\sigma$-algebras $\bigvee_{i=m}^{n}
S^{-i}\mathcal{A}$, $n\geq m$, where the latter consists of finite unions of cylinders  $_m[e_m,...,e_n]$.

Let $\phi_0$ be a finite non-negative measure on $\mathcal{A}_0$.
Let $\phi_m$ denote the measure on $\mathcal{A}_m$ given by
\[\phi_m:= \phi_0\circ S^m\]
for all $m\leq 0$.

In \cite{Wer3} the following dynamically defined outer measure on $\Sigma$ was introduced. It arose very naturally from the need of measuring some subsets of $\Sigma$ which depend on the whole past under circumstances where they can be covered only with infinitely many sets $A_m\in\mathcal{A}_m$  with known values $\phi_m(A_m)$, $m\leq 0$.
\begin{Definition}\label{dom}
 For $Q\subset\Sigma$, set
 \[\mathcal{C}(Q):=\left\{(A_m)_{m\leq 0}:A_m\in\mathcal{A}_m\
 \forall m\mbox{ and }Q\subset\bigcup\limits_{m\leq 0}A_m \right\}\]
 and
 \[\Phi(Q):=\inf\left\{\sum\limits_{m\leq0}\phi_m(A_m):(A_m)_{m\leq 0}\in\mathcal{C}(Q)\right\}.\]
\end{Definition}
It is quite straightforward  to check that $\Phi$ defines an outer measure on $\Sigma$ (e.g. see \cite{Wer3}). Moreover, it is not difficult to verify that $\Phi$ is exactly the outer measure used for the Carat\'{e}odory's construction of a measure  if  $\phi_m$'s satisfy  Kolmogorov's consistency condition (see Proposition \ref{kce}). In the following, it will be shown, in particular, that our generalization of Carat\'{e}odory's outer measure is a way for obtaining measures on product spaces from measures on subspaces which do not satisfy Kolmogorov's consistency condition, but are consecutive images of each other under the shift map.

An other observation which can be made at this point is that the definition is possible only by the Axiom of Choice.
However, this will not repel a scientifically minded reader because it will be shown below (Lemma 5) that a family of covering sets $(A_m)_{m\leq 0}$ of a Borel set can be chosen  in such a way that only finitely many of them are non-empty. Therefore, the definition of $\Phi$ can be legitimately called a construction.

This is important because it will be shown in Section 2 that $\Phi$ allows to obtain certain equilibrium states. The construction of equilibrium states for dynamical systems is one of the tasks of mathematical physics. It became a subject of a rigorous  mathematical analysis under the name of {\it Gibbs measures} \cite{K} \cite{R1} \cite{R2} \cite{S1} since the seminal works of Bogolubov and Hacet \cite{BH}, Dobrushin \cite{D}, Ruelle \cite{R} and Sinai \cite{S}, generalizing the {\it Boltzmann distribution}, which has been used in physics as a distribution which minimizes the {\it free energy} of certain systems. The main motivation for these efforts was the construction of physically  meaningful invariant measures for dynamical systems, which these systems asymptotically seek to achieve starting from measures about which we have some partial information. As far as the author is aware, the existence of equilibrium states is known only for upper semicontinuous energy functions  \cite{K}, but there are no universal way for a construction of them so far even for continuous functions (usually some stronger continuity conditions, such as {\it summability of variation}, are required). 

It will be shown in Section 2.1 that the construction of $\Phi$ allows to obtain equilibrium states
for some random dynamical systems  introduced in \cite{Wer1} as {\it contractive Markov systems}. They are a unifying generalization of {\it les cha\^{i}nes \`{a} liaisons compl\`{e}tes} \cite{DF}, {\it $g$-measures} \cite{Ke} and
{\it iterated function systems with place dependent probabilities}
\cite{BDEG}  which allows to formulate the results on  stability
of them in a language which is more systematic and natural for this
kind of questions. They are known to posses sometimes several stationary states \cite{BK},\cite{BHS}. Taking into the account the wide spectrum of applications of them in the modern science (e.g. \cite{BDX},\cite{Barnsley},\cite{BHMS},\cite{FHK},\cite{Sl},\cite{BLLC}) leaves no doubts that the questions on stability of them are no less important than curious. It was shown in \cite{Wer6} that the energy functions associated with such systems are not even upper semicontinuous in general (even in the case of the usual regularity conditions on the probability functions). Nevertheless, the existence of equilibrium states for them has been proved \cite{Wer6}. However, the task of the construction of equilibrium states for such energy functions from non-invariant initial measures remained open. All author's attempts to use the well know technics of Gibbs measures failed. Here we obtain them using the construction of $\Phi$ even under conditions which are much weaker than those assumed in \cite{Wer6}. The definition of $\Phi$ resembles some of the constructions of Gibbs measures \cite{S} and connects it with the general way in which measures are constructed in the Measure Theory \cite{B}. Moreover, it is not difficult to see that the definition of $\Phi$ can be easily extended to more general dynamical systems. 

Every theory is based just on a few examples. In almost every text book, Gibbs measures are illustrated with the help of finite Markov chains, which are just  trivial cases of contractive Markov systems. If an example does not fit into the theory, the latter needs to be modified. Every working  example is guiding in this situation and therefore should not be ignored.

Now, we will refine the definition to make it more easily applicable in some situations. Also, this will make the previous claims more obvious.

For each $m\leq 0$, let $\Delta_m$ denote the algebra consisting of all finite unions of cylinders of the form $_{m}[e_{m},e_{m+1}...,e_n]$, $e_i\in E$, $m\leq i\leq n$. Set
\begin{equation*}
\Xi(Q) := \left\{\left(C_m\right)_{m\leq 0}: C_m\in\Delta_m\mbox{ for all }m\leq 0\mbox { and }Q\subset\bigcup\limits_{m\leq 0} C_m\right\}
\end{equation*}
for all $Q\in\Sigma$.

Since each $\Delta_m$ generates the $\sigma$-algebra $\mathcal{A}_m$, the following lemma shows that
the definition of $\Phi$ is a generalization of the usual definition of an outer measure from the standard Measure Theory, which is used for the Carat\'{e}odory's method for obtaining a measure, as mentioned above.

\begin{Lemma}
  Let $Q\subset\Sigma$. Then
  \begin{equation*}
      \Phi(Q) = \inf\left\{\sum\limits_{m\leq 0}\phi_m(C_m): \left(C_m\right)_{m\leq 0}\in\Xi(Q)\right\}.
  \end{equation*}
\end{Lemma}

{\it Proof.}
Clearly, the left hand side of the claimed equation can not exceed it's right hand side.

On the other hand,
by the definition of $\Phi(Q) $, there exists a sequence $\left((A^n_m)_{m\leq 0}\right)_{n\in\mathbb{N}}\subset\mathcal{C}(B)$ such that
\begin{equation*}
    \sum\limits_{m\leq 0}\phi_m(A^n_m)\downarrow \Phi(Q)\mbox{ as }n\to\infty.
\end{equation*}
By the standard approximation theory, for every $n\in\mathbb{N}$ and $m\leq 0$ there exists $C^n_m\in\Delta_m$ such that
$ A^n_m\subset C^n_m$ and
\begin{equation*}
    \phi_m(C^n_m\setminus A^n_m)<\frac{2^m}{n}.
\end{equation*}
Therefore,
\begin{equation*}
\left|\sum\limits_{m\leq 0}\phi_m\left(C^n_m\right) - \Phi(Q)\right|\leq\frac{1}{n} + \left|\sum\limits_{m\leq 0}\phi_m\left(A^n_m\right) - \Phi(Q)\right|.
\end{equation*}
This implies the claim.
\hfill$\Box$

The class of the covering sets in the construction of $\Phi$ can be restricted even further. Set
\begin{equation*}
\dot\Xi(Q) := \left\{\left(B_m\right)_{m\leq 0}: B_m\in\Delta_m, B_m\cap B_n = \emptyset\ \forall m\neq n\mbox { and }Q\subset\bigcup\limits_{m\leq 0} B_m\right\}
\end{equation*}
for all $Q\subset\Sigma$.

\begin{Lemma}\label{dcl}
  Let $Q\subset\Sigma$. Then
  \begin{equation*}
      \Phi(Q) = \inf\left\{\sum\limits_{m\leq 0}\phi_m(B_m): \left(B_m\right)_{m\leq 0}\in\dot\Xi(Q)\right\}.
  \end{equation*}
\end{Lemma}

{\it Proof.}
Let $\dot\Phi(Q)$ denote the right hand side of the equation to be proved. Then obviously
\begin{equation*}
\dot\Phi(Q) \geq \Phi(Q).
\end{equation*}
Now, let $(C_m)_{m\leq 0}\in\Xi(Q)$. Set
\begin{equation*}
B_m:=C_m\setminus\left(C_{m+1}\cup...\cup C_0\right)\mbox{ for all }m\leq 0.
\end{equation*}
Then $(B_m)_{m\leq 0}\in\dot\Xi(Q)$ and $B_m\subset C_m$ for all $m\leq 0$. Hence
\begin{equation*}
    \sum\limits_{m\leq 0}\phi_m(C_m)\geq \sum\limits_{m\leq 0}\phi_m(B_m)\geq \dot\Phi(Q).
\end{equation*}
Therefore,
\begin{equation*}
\Phi(Q) \geq \dot\Phi(Q).
\end{equation*}
This concludes the proof.
\hfill$\Box$

Now, we give  a formal proof that $\Phi$ is a natural dynamical extension of the  Carat\'{e}odory's outer measure which contains the latter as the case with consistent measures $\phi_m$'s.

\begin{Proposition}\label{kce}
  Suppose that $\phi_{-1}$ is consistent with $\phi_0$, i.e. $\phi_{-1}(C ) = \phi_0(C)$ for all cylinder sets $C\in\mathcal{A}_0$.
  Then $\Phi$ coincides with Carat\'{e}odory's outer measure which extends the set function $\phi$ defined for every cylinder set $C\in\mathcal{A}_m$, $m\leq 0$, by $\phi(C) := \phi_m(C)$.
\end{Proposition}

{\it Proof.}
   Let $M$ denote Carat\'{e}odory's outer measure extending the set function $\phi$. Let $Q\subset\Sigma$ and $(A_m)_{m\leq 0}\in \Xi(Q)$. Since each $A_m$ can be written as a finite disjoint union of some cylinders $C_{1m},...,C_{n_mm}$ from $\mathcal{A}_m$,
   \begin{eqnarray*}
 \sum\limits_{m\leq 0}\phi_m(A_m) = \sum\limits_{m\leq 0}\sum\limits_{k=1}^{n_m}\phi\left(C_{km}\right)\geq M(Q).
\end{eqnarray*}
Hence,
\[\Phi(Q)\geq M(Q).\]
Now, let $(C_n)_{n\in\mathbb{N}}$ be a sequence of cylinders such that $Q\subset \bigcup_{n\in\mathbb{N}}C_n$. Let $m_1$ be the largest non-positive integer such that $C_1\in\mathcal{A}_{m_1}$. Set $B_{m_1}:=C_1$ and $B_m=\emptyset$ for all $m_1<m\leq 0$. Recursively, assuming that all $B_m$'s  are defined for all $m_n\leq m\leq 0$, choose $m_{n+1}<m_n$ such that $C_{n+1}\in \mathcal{A}_{m_{n+1}}$ and set $B_{m_{n+1}}:=C_{n+1}$ and $B_m = \emptyset$ for all $m_{n+1}<m<m_n$. Then
$(B_m)_{m\leq 0}\in \Xi(Q)$. Therefore,
\[\sum\limits_{n\in\mathbb{N}}\phi(C_n) = \sum\limits_{m\leq 0}\phi(B_m)=\sum\limits_{m\leq 0}\phi_m(B_m)\geq \Phi(Q).\]
Thus, \[M(Q)\geq\Phi(Q).\] This completes the proof.
\hfill$\Box$

The following Lemma is useful for verifying that $\Phi(\Sigma)>0$.
\begin{Lemma}\label{acl}
  Suppose $\phi'_0$ is a measure on $\mathcal{A}_0$ which is absolutely continuous with respect to $\phi_0$. Let $\Phi'$ denote the outer measure obtained from  $\phi'_0$ as in Definition \ref{dom}. Then for every $\epsilon>0$ there exists $\delta>0$ such that, for every $Q\subset\Sigma$, $\Phi'(Q)<\epsilon$ whenever $\Phi(Q)<\delta$.
\end{Lemma}

{\it Proof.} See the proof of Lemma 2 (ii) in \cite{Wer3}.
\hfill$\Box$

Next, we are going to investigate what happens to $\Phi$ under the action of the shift map. The action can be considered into two ways.
\begin{Definition}\label{simd}
 For $m\leq 0$, let $\Phi_{(m)}$ denote the outer measure $\Phi$ where $\phi_m$ is taken as the initial measure on $\mathcal{A}_0$, instead of $\phi_0$.
\end{Definition}

\begin{Lemma}\label{oms}
  Let $Q\subset\Sigma$. Then
  \begin{equation*}
      \Phi(Q) \leq \Phi(S^{-1}Q) \leq \Phi_{(-1)}(Q).
  \end{equation*}
\end{Lemma}

{\it Proof.}
Let $(A_m)_{m\leq 0}\in \Xi(Q)$. Then $(...,SA_{-1}, SA_0,\emptyset)\in\Xi(SQ)$. Therefore,
\begin{equation*}
   \Phi(SQ) \leq\sum\limits_{m\leq 0}\phi_{m-1}(SA_m).
\end{equation*}
Since   $\phi_{m-1}(SA_m) = \phi_{m}(A_m)$ for all $m\leq 0$,
\begin{equation*}
   \Phi(SQ) \leq\sum\limits_{m\leq 0}\phi_{m}(A_m).
\end{equation*}
This implies
\begin{equation*}
      \Phi(SQ) \leq \Phi(Q),
  \end{equation*}
  which is equivalent to the first inequality.

  On the other hand, since $(S^{-1}A_m)_{m\leq 0}\in \Xi(S^{-1}Q)$,
  \[\Phi\left(S^{-1}Q\right)\leq\sum\limits_{m\leq 0}\phi_{m}\left(S^{-1}A_m\right).\]
  This gives
  \[\Phi(S^{-1}Q)\leq \Phi_{(-1)}(Q).\]

\hfill$\Box$

Obviously $\Phi$ is finite since $\phi_0$ is finite. Therefore, by Lemma \ref{oms}, we can make the following definition.
\begin{Definition}\label{simd}
 For $Q\subset\Sigma$, set
   \[\Phi^*(Q) := \lim\limits_{m\to-\infty}\Phi_{(m)}\left(Q\right).\]
   Obviously, $\Phi^*$ is an outer measure. By Lemma \ref{oms},  $\Phi\leq \Phi^*$.
\end{Definition}

Next theorem bears the fruits of the construction.

\begin{Theorem}\label{sibm}
   The restrictions of $\Phi$ and $\Phi^*$ on the Borel $\sigma$-algebra are shift invariant measures.
   \end{Theorem}
   {\it Proof.}
   We start with $\Phi$.  By Lemma \ref{oms}, we can define
   \[\bar\Phi(Q) := \lim\limits_{m\to-\infty}\Phi\left(S^mQ\right)\]
   for all $Q\subset\Sigma$.
   It is easy to check that $\bar\Phi$ is an outer measure on $\Sigma$. Also, it is  clear that it is shift invariant.
   First, we show that the restriction of $\bar\Phi$ on the Borel $\sigma$-algebra on $\Sigma$ is a measure.
   We will demonstrate it using Carat\'{e}odory's method, i.e by showing that the $\sigma$-algebra of all $\bar\Phi$-measurable  sets contains all cylinder sets.
   That is we need to show that
  \begin{equation}\label{mc}
       \bar\Phi(Q)\geq \bar\Phi(Q\cap C) + \bar\Phi(Q\setminus C)
  \end{equation}
 for all cylinder sets $C$ and all $Q\subset\Sigma$.

 So, let $C$ be a cylinder set and $Q$ a subset of $\Sigma$. Then there exists $n\geq 0$ such that $S^{-i}C\in\Delta_0$ for all $i\geq n$ and hence $S^{-i}C\in\Delta_m$ for all $i\geq n$ and $m\leq 0$. Let $i\geq n$ and $(A_m)_{m\leq 0}\in\mathcal{C}(S^{-i}Q)$. Then
 \begin{equation*}
     \sum\limits_{m\leq 0}\phi_{m}\left(A_m\right) = \sum\limits_{m\leq 0}\phi_{m}\left(A_m\cap S^{-i}C\right) + \sum\limits_{m\leq 0}\phi_{m}\left(A_m\setminus S^{-i}C\right).
 \end{equation*}
   Since $(A_m\cap S^{-i}C)_{m\leq 0}\in\mathcal{C}(S^{-i}(Q\cap C))$ and $(A_m\setminus S^{-i}C)_{m\leq 0}\in\mathcal{C}(S^{-i}(Q\setminus C))$,
   \begin{equation*}
      \sum\limits_{m\leq 0}\phi_{m}\left(A_m\right) \geq \Phi\left(S^{-i}(Q\cap C)\right) + \Phi\left(S^{-i}(Q\setminus C)\right).
   \end{equation*}
  Hence
  \begin{equation*}
     \Phi\left(S^{-i}Q\right) \geq \Phi\left(S^{-i}(Q\cap C)\right) + \Phi\left(S^{-i}(Q\setminus C)\right).
   \end{equation*}
   Taking the limit gives (\ref{mc}).

Let $B\subset\Sigma$ be a Borel set. The first part of the claim will
follow, if we show that $\Phi(B)$ = $\bar\Phi(B)$.  By Lemma
\ref{oms},
        \[\Phi(B) \leq \bar\Phi(B).\]
        Let $B'\subset\Sigma$ be a Borel set. Since $\bar\Phi$ is a Borel measure and $\Phi$ is an outer measure with $\Phi(\Sigma) = \bar\Phi(\Sigma)$, by Lemma \ref{oms},
         \begin{eqnarray*}
           \bar\Phi(\Sigma\setminus B') = \bar\Phi(\Sigma) - \bar\Phi(B')
           \leq\Phi(\Sigma) - \Phi(B')
           \leq\Phi(\Sigma\setminus B').
       \end{eqnarray*}
   So, for $B':= \Sigma\setminus B$, we get
   \[\Phi(B) \geq \bar\Phi(B),\]
   as desired. 
   
   It is possible to deduce the second part of the claim
   from the first. Here we give a proof of it which is
   independent from the proof of the first. To show that $\Phi^*$ is a shift invariant Borel measure, observe that, by Lemma \ref{oms},
   \[ \Phi_{(-i)}(Q)\leq \Phi_{(-i)}(S^{-1}Q)\leq \Phi_{(-i-1)}(Q)\]
   for all $i\in\mathbb{N}$. Taking the limit gives
   \[{\Phi^*}(S^{-1}Q) = {\Phi^*}(Q).\]
   Now, we show in  the same way that ${\Phi^*}$ is a Borel measure. Let $(A_m)_{m\leq 0}\in\mathcal{C}(Q)$.
   Observer that $(S^{-i}(A_m\cap C))_{m\leq 0}\in\mathcal{C}(S^{-i}(Q\cap C))$ and $(S^{-i}(A_m\setminus C))_{m\leq 0}\in\mathcal{C}(S^{-i}(Q\setminus C))$ for all $i\geq n$. Therefore,
  \begin{eqnarray*}
     \sum\limits_{m\leq 0}\phi_{m - 2i}\left(A_m\right)
     &=& \sum\limits_{m\leq 0}\phi_{m-i}\left(S^{-i}A_m\right)\\
     &=& \sum\limits_{m\leq 0}\phi_{m-i}\left(S^{-i}(A_m\cap C)\right) + \sum\limits_{m\leq 0}\phi_{m-i}\left(S^{-i}(A_m\setminus C)\right)\\
     &\geq&\Phi_{(-i)}\left(S^{-i}(Q\cap C)\right) + \Phi_{(-i)}\left(S^{-i}(Q\setminus C)\right)\\
     &\geq&\Phi_{(-i)}\left(Q\cap C\right) + \Phi_{(-i)}\left(Q\setminus C\right)
 \end{eqnarray*}
   for all $i\geq n$. Hence,
   \[\Phi_{(-2i)}\left(Q\right)\geq\Phi_{(-i)}\left(Q\cap C\right) + \Phi_{(-i)}\left(Q\setminus C\right).\] Taking the limit gives
   \[\Phi^*(Q)\geq \Phi^*(Q\cap C) + \Phi^*(Q\setminus C).\]
   This completes the proof.
   \hfill$\Box$

   In the following, we will use the same notation for measures obtained in  Theorem \ref{sibm} as for outer measures $\Phi$ and $\Phi^*$ if no confusion is possible.

 At this point, it is easy to see that the family of the covering sets of Borel sets in the definition of $\Phi$ can be restricted further.
 Set
 \[\dot\zeta (B):=\{(A_m)_{m\leq 0}\in\dot\Xi(B):\ \mbox{ at most finitely many }A_m\neq\emptyset\}\]
 for all $B\subset\Sigma$. Choosing elements from finitely many $\sigma$-algebras does not requite The Axiom of Choice. The next lemma brings $\Phi$ back to the constructive world.
   \begin{Lemma}\label{fcl}
        Let $B\subset\Sigma$ be Borel. Than
        \[\Phi(B)=\inf\limits_{(A_m)_{m\leq 0}\in\dot\zeta (B)}\sum\limits_{m\leq 0}\phi_m(A_m).\]
    \end{Lemma}
    {\it Proof.} First let $C\subset\Sigma$ be compact. Obviously,
     \[\Phi(C)\leq\inf\limits_{(A_m)_{m\leq 0}\in\dot\zeta (C)}\sum\limits_{m\leq 0}\phi_m(A_m).\]
     On the other hand, by the compactness of $C$, for every $(A_m)_{m\leq 0}\in\dot\Xi(C)$ there exists $(A'_m)_{m\leq 0}\in\dot\zeta(C)$ such that
     \[\sum\limits_{m\leq 0}\phi_m(A_m)\geq\sum\limits_{m\leq 0}\phi_m(A'_m)\geq\inf\limits_{(A'_m)_{m\leq 0}\in\dot\zeta (C)}\sum\limits_{m\leq 0}\phi_m(A'_m).\]
     Hence
     \[\Phi(C)\geq\inf\limits_{(A_m)_{m\leq 0}\in\dot\zeta (C)}\sum\limits_{m\leq 0}\phi_m(A_m).\]
    Since the class of compact subsets contains the family of all cylinder sets, the claim follows by Theorem \ref{sibm}.
    \hfill$\Box$

     It has been shown in Theorem \ref{sibm} that  the definition of $\Phi$ let's to obtain a nice shift invariant measure on $\Sigma$. The next example demonstrates that it works not always.

   \begin{Example}\label{te}
      Let $\Sigma := \{0,1\}^{\mathbb{Z}}$ and $\sigma'\in\Sigma$ given by
       \begin{equation*}
    \sigma'_i=\left\{\begin{array}{cc} 0&  \mbox{if }i \mbox{ is even }\\
    1& \mbox{ otherwise }
     \end{array}\right.
     \end{equation*}
     for all $i\in\mathbb{Z}$. Let $\phi_0$ be the probability measure on $\mathcal{A}_0$ given by
      \begin{equation*}
    \phi_0(A)=\left\{\begin{array}{cc} 1&  \mbox{if }\sigma'\in A\\
    0& \mbox{ otherwise }
     \end{array}\right.
     \end{equation*}
     for all $A\in\mathcal{A}_0$. Then for $(A_m)_{m\leq 0}\in\dot\Xi(\Sigma)$ given by $A_0:=\ _0[1]$, $A_{-1}:=\ _0[0]$ and $A_m := \emptyset$ for all $m\leq -2$, $\sum_{m\leq 0}\phi_m(A_m) = 0$. Hence $\Phi(\Sigma) = 0$. Similarly, by swapping  $_0[0]$ and $_0[1]$ if necessary, one sees that each $\Phi_{(-n)}(\Sigma) = 0$. Thus $\Phi^*(\Sigma) = 0$ also.
   \end{Example}

     However, from \cite{Wer3} we know examples where $\Phi$ gives useful non-zero measures.  It is natural to hope that, in case $\Phi(\Sigma)> 0$, it is a probability measure if $\phi_0$ is one.  The following proposition shows in particular that it is not true in general .

   \begin{Proposition}\label{npr}
        Suppose $\phi_0$ is a probability measure. Then\\
        (i) $\Phi(\Sigma) \leq 1$,\\
        (ii) $\left|\phi_m(A) - \phi_0(A)\right|\leq 1- \Phi(\Sigma)$ for all $A\in\mathcal{A}_0$ and $m\leq 0$.\\
       (iii)  The following are equivalent:\\
                a) $\Phi(\Sigma) = 1$,\\
                b) $\phi_{0}(S^{-1}A) = \phi_0(A)$ for all $A\in\mathcal{A}_0$,\\
                c) $\Phi$  uniquely extends $(\phi_m)_{m\leq 0}$ on the Borel $\sigma$-algebra.
    \end{Proposition}
    {\it Proof.}
    (i) It is clear that $\Phi(\Sigma) \leq 1$, since $(...\emptyset, \emptyset, \Sigma)\in\mathcal{C}(\Sigma)$ and $\phi_0$ is a probability measure.

      (ii) By the definition of $\Phi$ and Theorem \ref{sibm}, for every $A\in\mathcal{A}_0$ and $m\leq 0$,
      \[1-\phi_m(A) = \phi_m(\Sigma\setminus A)\geq \Phi(\Sigma\setminus A) = \Phi(\Sigma) - \Phi(A)\geq \Phi(\Sigma) - \phi_0(A).\]
      Hence \[\phi_m(A) - \phi_0(A)\leq 1- \Phi(\Sigma).\] Applying this inequality to $\Sigma\setminus A$ gives (ii).

      (iii) The implication form a)  to b) follows immediately from (ii). The implication from b) to c) follows from
      Proposition \ref{kce} and Kolmogorov's Consistency Theorem. The implication from c) to a) is obvious.
        \hfill$\Box$

        By Proposition \ref{npr}, a normalization of $\Phi$ is in general necessary if $\Phi(\Sigma)>0$, which is a typical situation in statistical mechanics. Non-typical in this approach is that we don't need to require some convergence in so-called 'thermodynamic limit' to obtain a measure on the Borel $\sigma$-algebra.

   We conclude this section by giving some sufficient conditions on $\phi_0$ for the positivity of $\Phi(\Sigma)$.

 \begin{Proposition}\label{pc}
    (i) Let $\nu$ be a positive Borel measure on $\Sigma$.    Let $\phi'_m$ denote the absolutely continuous part of the Lebesgue decomposition of $\phi_m$ with respect to $\nu$. Suppose there exists a Borel-measurable function $f$ such that $d\phi'_m/d\nu\geq f$ for all $m\leq 0$ and $\int f\ d\nu>0$.  Then $\Phi(\Sigma)>0$.\\
    (ii) Suppose there exists a positive shift invariant measure on $\mathcal{A}_0$ which is absolutely continuous with respect to $\phi_0$. Then $\Phi(\Sigma)>0$.
    \end{Proposition}
     {\it Proof.}
     (i) Observe that
    \[\sum\limits_{m\leq 0}\phi_m(A_m)\geq\sum\limits_{m\leq 0}\int\limits_{A_m}f\ d\nu \geq \int f\ d\nu>0\]
    for all $(A_m)_{m\leq 0}\in\mathcal{C}(\Sigma)$. Therefore, the claim follows.

    (ii) The claim follows by Lemma \ref{acl} and Proposition \ref{npr} (iii).
    
 \hfill$\Box$

Next section provides in particular some physically meaningful examples where $\Phi(\Sigma)>0$.

     \section{Equilibrium states}

     In this section, we intend to show that construction of $\Phi$ allows to obtain equilibrium states for some energy function $u: \Sigma\longrightarrow[-\infty,0]$.

   \begin{Definition}
   Let $P_S(\Sigma)$ denote the space of all shift invariant Borel probability measures on $\Sigma$ and $h_{\Lambda}(S)$   be the {\it Shannon-Kolmogorov-Sinai entropy} of $S$ with respect to  $\Lambda\in P_S(\Sigma)$. $\Lambda_0\in P_S(\Sigma)$ is said to be an {\it equilibrium state} for $u$ iff
         \begin{equation*}
      h_{\Lambda_0}(S) + \int u d\Lambda_0 = \sup\limits_{\Lambda\in P_S(\Sigma)}\left\{ h_{\Lambda}(S) + \int u d\Lambda\right\}.
   \end{equation*}
  The physical interpretation of it is that the equilibrium state is a state which minimizes the {\it free energy} of the system.
   \end{Definition}
   To proceed towards our goal, we need to split each $\sigma$-algebra $\mathcal{A}_m$, $m\leq 0$, into two pieces, the one which depends on the past and the other which depends on the future.

   \begin{Definition}\label{psa}
      Let $G$ denote the $\sigma$-algebra on $\Sigma$ which is generated by cylinders  of the form $_1[e_1,...,e_n]$, $e_i\in E$, $1\leq i\leq n$, $n\in\mathbb{N}$. For $m\leq 0$, let $\mathcal{F}_m$ denote the finite $\sigma$-algebra on $\Sigma$ generated by cylinders of the form $_m[e_m,...,e_0]$, $e_i\in E$, $m\leq i\leq  0$. Finally, let $\mathcal{F}$ denote the $\sigma$-algebra generated by $\bigcup_{m\leq 0}\mathcal{F}_m$.
   \end{Definition}

    Recall that by Kolmogorov-Sinai Theorem,
    \begin{equation}\label{KSE}
        h_{\Lambda}(S) = -\sum\limits_{e\in E}\int E_{\Lambda}\left(1_{_1[e]}|\mathcal{F}\right)\log E_{\Lambda}\left(1_{_1[e]}|\mathcal{F}\right) d\Lambda,
    \end{equation}
where $E_{\Lambda}\left(1_{_1[e]}|\mathcal{F}\right)$ denotes the conditional expectation of the indicator function $1_{_1[e]}$ conditioned on $\mathcal{F}$ with respect to $\Lambda$. $h_{\Lambda}(S)$ is interpreted as a measure of uncertainty of observing the next symbol of process $\Lambda$ given its past. This is the reason for the split on the past and the future.

From (\ref{KSE}) follows the desire to be able to compute the conditional expectations with respect to $\Phi$, which might help to verify the equilibrium state property. This seems to require an extension of the standard theory of Martingales to adapted random variables with different underlying probability measures. Such an extension is done in the next theorem, which is a simple consequence of the construction of $\Phi$. This can be compared to the result in \cite{LR}, where $\phi_m$'s are assumed to be defined on the limiting $\sigma$-algebra and converge in some sense.

     \begin{Theorem}\label{gmt}
   Let $f:\Sigma\longrightarrow\mathbb{R}$ be $G$-measurable and bounded. Suppose $\Phi(\Sigma) > 0$.
   Then
       \[E_{\phi_{m}}\left(f|{\mathcal{F}_m}\right) \to E_{\Phi}\left(f|{\mathcal{F}}\right)\ \ \ \Phi\mbox{-a.e.} .\]
   \end{Theorem}
   {\it Proof.}
   Though the theorem appears to be a generalization of Doob's Martingale Theorem, the proof of it reduces to the truth of latter (just as the result in \cite{LR} does).

   Let $m\leq 0$. Since $\Phi|_{\mathcal{F}_m}\leq \phi_m|_{\mathcal{F}_m}$, by Radon-Nikodym Theorem, there exists $\xi_m:\Sigma\longrightarrow\mathbb{R}$ $\mathcal{F}_m$-measurable such that $\Phi(A) = \int_{A}\xi_md\phi_m$ for all $A\in\mathcal{F}_m$. Therefore, by the pool-out-property of the conditional expectation,
   \begin{eqnarray*}
      \int\limits_{A}E_{\phi_m}\left(f|{\mathcal{F}_m}\right)d\Phi&=&\int\limits_{A}\xi_mfd\phi_m
      =\int\limits_{A}fd\Phi
  \end{eqnarray*}
  for all $A\in\mathcal{F}_m$.
 Thus
 \begin{equation}\label{coee}
   E_{\phi_m}\left(f|{\mathcal{F}_m}\right) = E_{\Phi}\left(f|{\mathcal{F}_m}\right)\ \ \ \Phi\mbox{-a.e.}
\end{equation}
for all $m\leq 0$. Since the normalization of $\Phi$ dose not alter $E_{\Phi}\left(f|{\mathcal{F}_m}\right)$, by Doob's Martingale Theorem we conclude that
  \[E_{\phi_{m}}\left(f|{\mathcal{F}_m}\right) \to E_{\Phi}\left(f|{\mathcal{F}}\right)\ \ \ \Phi\mbox{-a.e.} .\]
       \hfill$\Box$

       \subsection{Equilibrium states for random dynamical systems}

       Now, we are going to apply the theory developed in this paper to some random dynamical systems introduced in \cite{Wer1} as {\it Markov systems}.
 It is known that the asymptotic behavior of contractive Markov systems  is similar \cite{Wer1}, to some extent, to the trivial case of  finite
{\it Markov chains}. However, the question on the necessary and
sufficient condition for the uniqueness of the stationary state for
such systems remains open already for more than seventy years. A
reason for that might be the lack of suitable mathematical tools.
This paper provides some new tools.
       They allow us to obtain new useful results about Markov system.
       On the other hand, this gives an example of non-zero measures $\Phi$ obtained from measures on sub-$\sigma$-algebras  which do not satisfy Kolmogorov's consistency condition. 

It was pointed out by an anonymous referee that it might be appropriate to cite in this paper the work by A. Lasota and J. Yorke \cite{LY} where a general method, called the lower bound technique, useful in providing criteria for a stability of such systems is presented.

   Let $\left(K_{i(e)},w_e,p_e\right)_{e\in E}$ be a Markov system \cite{Wer1}, i.e. $K_1,...,K_N$ is a partition of a complete metric space $(K,d)$ into non-empty Borel subsets,  $i:E\longrightarrow \{1,...,N\}$ surjective and $t:E\longrightarrow \{1,...,N\}$ such that, for every $e\in E$, $w_e: K_{i(e)}\longrightarrow K_{t(e)}$ and $p_e: K_{i(e)}\longrightarrow (0,1]$ such that $\sum_{e\in E, i(e) = i}p_e(x) = 1$ for all $x\in K_i$, $i\in\{1,...,N\}$. We can consider each $p_e$ to be extended on $K$ by zero and each $w_e$ to be extended on $K$ arbitrarily. We assume that each $p_e|_{K_{i(e)}}$ and $w_e|_{K_{i(e)}}$ is uniformly continuous, where notation $|_{A}$ means the restriction on a set $A$.

A Markov system is called {\it contractive} iff there exists $0<a<1$ such that
 \begin{equation}\label{cac}
 \sum\limits_{e\in E}p_e(x)d(w_ex,w_ey)<ad(x,y)\\\\\mbox{ for all }x,y\in K_i, i=1,...,N.
 \end{equation}
Let $\mathcal{B}(K)$ denote the Borel $\sigma$-algebra  and $P(K)$ the set of all Borel probability measures on $K$. Let $U$ be the Markov operator acting on real-valued functions $f$ on $K$ by
\[Uf = \sum\limits_{e\in E}p_ef\circ w_e\] and
$U^*$ be its adjoint operator acting on Borel probability measures $\nu$ on $K$ by $U^*\nu(B) = \int U(1_B)d\nu$ for all $B\in \mathcal{B}(K)$.

   Let $x\in K$.  For each integer $m\leq 1$, let $P_x^m$ be the probability measure on the
$\sigma$-algebra $\mathcal{A}_m$ given by
\[P^m_x(\ _{m}[e_{m},...,e_n])=p_{e_{m}}(x)p_{e_{m+1}}(w_{e_{m}}(x))...p_{e_n}(w_{e_{n-1}}\circ...\circ
w_{e_{m}}(x))\] for all cylinder sets $_{m}[e_{m},...,e_n]$,
$n\geq{m}$. It has been shown in \cite{Wer3} Lemma 1 that the function $x\longmapsto P^m_x(A)$ is Borel measurable for all $A\in\mathcal{A}_m$ (for that only the Borel measurability of $p_e$'s and $w_e$'s is needed).
   \begin{Definition}
 For $\nu\in P(K)$, let $\phi_m(\nu)$ be the probability measure  on $\mathcal{A}_m$ given by
 \[\phi_m(\nu)(A):=\int P_x^m(A)d\nu(x)\\\ \mbox{ for all } A\in\mathcal{A}_m.\]
 \end{Definition}

Observer that, for each $i\geq 0$, $\phi_{m}({U^*}^i\nu)$ and $\phi_m(\nu)\circ S^{-i}$ also define measures on $\mathcal{A}_m$ for all $m\leq 0$. The following lemma states their relations to  $\phi_{m}(\nu)$.

\begin{Lemma}\label{shm}
 Let $m\leq 0$ and $\nu\in P(K)$. Then

 (i) $ \phi_{m-1}(\nu)(Q) = \phi_m(U^*\nu) (Q) = \phi_m(\nu)\left(S^{-1}Q\right)$ for all $Q\in\mathcal{A}_m$,

(ii) $\phi_{m-1}(\nu) (Q) = \phi_m(\nu)\left(S^{-1}Q\right)$ for all $Q\in\mathcal{A}_{m-1}.$
\end{Lemma}
{\it Proof.}
(i) We show only the second equation, the proof of the first is the same.
It is sufficient to check that the measures agree on all cylinders generating $ \mathcal{A}_m$.
\begin{eqnarray*}
   \phi_m(\nu)\left({S^{-1}}  _m[e_1,...e_n]\right) &=& \phi_m(\nu)\left(  _{m+1}[e_1,...e_n]\right)\\
   &=&\int\sum\limits_{e\in E} p_{e}(x) P^m_{w_e(x)}\left( _m[e_1,...e_n]\right) d\nu(x)\\
   &=&\int P^m_{x}\left( _m[e_1,...e_n]\right) dU^*\nu(x)\\
   &=& \phi_m(U^*\nu)\left(  _m[e_1,...e_n]\right).
\end{eqnarray*}
For (ii), observe that
\[ \phi_m(\nu)\left({S^{-1}}  _{m-1}[e_1,...e_n]\right) =  \phi_m(\nu)\left(  _m[e_1,...e_n]\right) =  \phi_{m-1}(\nu)\left(  _{m-1}[e_1,...e_n]\right).\]
\hfill$\Box$

\begin{Definition}
For $\nu\in P(K)$, let $\Phi(\nu)$ and $\Phi^*(\nu)$ be the outer measures and measures as defined in (\ref{dom}) and (\ref{simd}) with  $\phi_m(\nu)$'s standing for $\phi_m$'s.
\end{Definition}

\begin{Remark}\label{cr}
Observe that, by Lemma \ref{shm} (i),  measures $\phi_m(\nu)$ satisfy Kolmogorov's consistency condition if $\nu$ is an invariant initial distribution of the Markov system. In this case, outer measure $\Phi(\nu)$ is the usual outer measure used for Carat\'{e}odory's construction of a measure.  In general, $\phi_m(\nu)$'s have exactly the properties of $\phi_m$'s from Section 1. In particular,  outer measure $\Phi_{(-i)}(\nu)$ is nothing else but $\Phi({U^*}^i\nu)$ for all $i\geq 0$.
\end{Remark}

The following simple lemma states some properties of the map $\nu\longmapsto\Phi(\nu)$.
\begin{Lemma}\label{pomm}
 Let  $\nu_1,\nu_2\in P(K)$. Then

 (i) $\Phi(\nu_1)\ll\Phi(\nu_2)$ if $\nu_1\ll\nu_2$, where $\ll$ denotes the absolute continuity relation.

(ii) For $0\leq\alpha\leq 1$,
\[\Phi(\alpha\nu_1+(1-\alpha)\nu_2)\geq\alpha\Phi(\nu_1)+(1-\alpha)\Phi(\nu_2). \]
\end{Lemma}
{\it Proof.} (i) Observe that $\phi_0(\nu_1)\ll\phi_0(\nu_2)$. Therefore, the claim follows by Lemma \ref{acl}.
(ii) is a direct consequence of the supperadditivity of the infimum.
\hfill$\Box$

We will use the following initial distributions for the Markov system.
\begin{Definition}
Fix $x_i\in K_i$ for all $i=1,...,N$ and set
\begin{equation*}
\nu_0 := \frac{1}{N}\sum\limits_{i=1}^N\delta_{x_i},
\end{equation*}
where $\delta_{x_i}$ denotes the Dirac probability measure concentrated at $x_i$. Let $\emptyset\neq S\subset\{1,...,N\}$ and set
 \[\nu'_0 :=  \frac{1}{|S|}\sum\limits_{i\in S}\delta_{x_i},\]
 where $|S|$ denotes the size of $S$.
\end{Definition}

\begin{Proposition}\label{omrds}
        (i) For $\nu\in P(K)$, the following are equivalent:\\
        a) $\Phi(\nu)(\Sigma)=1$,\\
        b) $U^*\nu = \nu$,\\
        c) $\Phi(\nu)$ uniquely extends $\phi_m(\nu)$'s on the Borel $\sigma$-algebra.\\
        (ii)
        Suppose there exists $\mu\in P(K)$ such that  $U^*\mu = \mu$ and  $P^0_x<<P^0_y$ for all $x, y\in K_i$ and $i\in S:=\{j:\mu(K_j)>0\}$.  Then
        $\Phi(\nu'_0)(\Sigma)>0.$
    \end{Proposition}
    {\it Proof.}
     (i) The claim follows from Lemma \ref{shm} and Proposition \ref{npr}.

     (ii)
      Clearly, it follows from the hypothesis that $P^0_x<<\phi_0(\nu'_0)$ for all $x\in K_i$ and $i\in S$. Therefore, $\phi_0(\mu)<<\phi_0(\nu'_0)$. Hence, by Lemma \ref{pomm}(i),  $\Phi(\mu)<<\Phi(\nu'_0)$.
      Since, by (i), $\Phi(\mu)(\Sigma) > 0$, the claim follows.
    \hfill$\Box$

   \begin{Example}\label{cr}
   It has been shown in \cite{HS} that there exists an invariant initial Borel probability distribution $\mu$ for a contractive Markov system on a Polish space if $(K_i)_{i=1}^N$ form an open partition of $K$. The condition of equivalence of measures $P^0_x$ and $P^0_y$ for all $x, y\in K_i$, $i=1,...,N$, is satisfied e.g. if probability functions $p_e|_{K_{i(e)}}$, $e\in E$, of a contractive Markov system have a {\it square summable variation} and are bounded away from zero \cite{Wer7}. By Proposition \ref{omrds} (ii), $\Phi(\nu_0)(\Sigma)>0$ for such systems. As far as the author is aware, the first result on the equivalence of measures $P^0_x$ and $P^0_y$ was obtained by J. H. Elton in \cite{E}, in the case when the partition has the single atom and the probability functions have a {\it summable variation} (Dini continuous) and are bounded away from zero.
\end{Example}

      \begin{Definition}
Set
\begin{equation*}
\Sigma_G :=\left\{\sigma\in\Sigma : i(\sigma_{n+1}) = t(\sigma_n)\mbox{ for all }n\in\mathbb{Z}\right\}
\end{equation*}
({\it subshift of finite type} associated with the Markov system) and
\begin{equation*}
D:=\left\{\sigma\in\Sigma_G : \lim\limits_{m\to-\infty}w_{\sigma_0}\circ ... \circ w_{\sigma_m}\left(x_{i(\sigma_m)}\right)\mbox{ exists }\right\}.
\end{equation*}
$D$ is a $\mathcal{F}$-measurable, e.g. one can see it quickly as
follows. Let's abbreviate $w^0_m(\sigma):=w_{\sigma_0}\circ ...
\circ w_{\sigma_m}(x_{i(\sigma_m)})$ for all $\sigma\in\Sigma$.
Clearly, $D=\{\sigma\in\Sigma_G:(w^0_m(\sigma))_{m\leq 0}\mbox{ is
Cauchy }\}$, by completeness of $K$. For each $m\leq 0$ and $n>0$
set $Q_{mn}:=\bigcap_{k\leq
0}\{\sigma\in\Sigma_G:d(w^0_m(\sigma),w^0_{m+k}(\sigma))<1/n\}$,
which are obviously $\mathcal{F}$-measurable sets. Then
$D=\bigcap_{n\in\mathbb{N}}\bigcup_{m\leq 0}Q_{mn}$ and therefore it
is $\mathcal{F}$-measurable.
\end{Definition}

   \begin{Proposition}\label{cmd}  Suppose the Markov system is contractive.  Then\\
   (i) $ \Phi(\nu'_0)(\Sigma\setminus D) = 0$ and\\
   (ii) $D\subset S^{-1}D$.
   \end{Proposition}
   {\it Proof.}
   (i) It was demonstrated in \cite{Wer3} that
   \begin{equation*}
       \Phi(\nu_0)\left(\Sigma\setminus D\right) = 0,
   \end{equation*}
   which is a direct consequence of the contractiveness condition (\ref{cac}) (no continuity of $p_e|_{K_{i(e)}}$'s or  $w_e|_{K_{i(e)}}$'s is required for that). Since $\nu'_0\ll\nu_0$, the claim follows by Lemma \ref{pomm} (i).

 (ii) Let $\bar K_{i(e)}$ denote the closure of $K_{i(e)}$, $e\in E$.
Since each $w_e|_{K_{i(e)}}$ is uniformly continuous, there exists a unique continuous extension of $w_e$ on $\bar K_{i(e)}$  (e.g. Theorem 2, p. 190 in \cite{Bou}), which we will denote with $\bar w_e$.  Let $\sigma\in D$, then there exists $y\in\bar K_{t(\sigma_0)}$ such that $y = \lim_{m\to-\infty}w_{\sigma_0}\circ ... \circ w_{\sigma_m}\left(x_{i(\sigma_m)}\right)$. Hence, $\bar w_{\sigma_1}(y) = \lim_{m\to-\infty}w_{\sigma_1}\circ w_{\sigma_0}\circ ... \circ w_{\sigma_m}\left(x_{i(\sigma_m)}\right)$. Therefore $S(\sigma)\in D$. That is $\sigma\in S^{-1}D$.
  \hfill$\Box$

   \begin{Definition}
    Set
       \begin{eqnarray*} \label{cm}
    F:\Sigma\longrightarrow K\nonumber
     \end{eqnarray*}
 by  \begin{equation*}
    F(\sigma):=\left\{\begin{array}{cc}
    \lim\limits_{m\to-\infty}w_{\sigma_0}\circ w_{\sigma_{-1}}\circ...\circ w_{\sigma_{m}}(x_{i(\sigma_{m})})&  \mbox{if }\sigma\in D\\
     x_{t(\sigma_0)}& \mbox{ otherwise, }
     \end{array}\right.
     \end{equation*}
 We call $F$ {\it the coding map} of the Markov system. It is
 obviously $\mathcal{F}$-Borel-measurable, as it is the pointwise limit of $\mathcal{F}$-Borel-measurable
 functions $F_m$ given by $F_m(\sigma):= w^0_m(\sigma)$ if $\sigma\in D$ and
 $F_m(\sigma)= x_{t(\sigma_0)}$ otherwise for all $m\leq 0$.
\end{Definition}
  Note that $F$ might be meaningful not only for contractive Markov systems. For example, if normalized $\Phi(\nu'_0)$ is ergodic, then  $\Phi(\nu'_0)(D) = 1$ or $\Phi(\nu'_0)(D) = 0$ by the shift invariance of $D$ (Proposition \ref{cmd} (ii)).

\begin{Lemma}\label{rpl}
   Suppose the Markov system is contractive.  Then $F\left(\Phi(\nu'_0)\right)$ is a Radon measure.
 \end{Lemma}
   {\it Proof.} Obviously, $F\left(\Phi(\nu'_0)\right)$ is a Borel measure, since $F$ is in particular Borel-Borel-measurable.
   Furthermore, $F\left(\Phi(\nu'_0)\right)$ is regular, since $K$ is a metric space (e.g. Theorem 7.1.7, p.70, Vol. II in \cite{B}). So, the claim will follow if we show that $F\left(\Phi(\nu'_0)\right)$ is tight, since an intersection of a compact and a closed set in $K$ is compact.

   Let $\epsilon>0$. By Lemma 3 (iii) in \cite{Wer3}, there exist  a compact set $Q\subset  \Sigma$ with $\Phi(\nu_0)(\Sigma\setminus Q)< \epsilon$ such that  $F|_{Q}$ is continuous. Set $C:=F|_{Q}(Q)$. Then $C$ is compact and
   \begin{eqnarray*}
   F\left(\Phi(\nu_0)\right) (K\setminus C)&=&\Phi(\nu_0)\left(\Sigma\setminus F^{-1}(C)\right)\\
   &\leq&\Phi(\nu_0)\left(\Sigma\setminus (F|_{Q})^{-1}(C)\right)\\
   &\leq&\Phi(\nu_0)\left(\Sigma\setminus Q\right)\\
   &<&\epsilon.
\end{eqnarray*}
 Hence $F\left(\Phi(\nu_0)\right)$ is tight. By Lemma \ref{pomm} (i), $F\left(\Phi(\nu'_0)\right)$ is absolutely continuous with respect to $F\left(\Phi(\nu_0)\right)$, therefore $F\left(\Phi(\nu'_0)\right)$ is also tight.
  \hfill$\Box$

   \begin{Definition}
       Set
      \begin{equation*}
    u(\sigma):=\left\{\begin{array}{cc}
    \log  p_{\sigma_1}\circ F(\sigma)&  \mbox{if }\sigma\in D\\
    -\infty& \mbox{ otherwise, }
     \end{array}\right.
     \end{equation*}
     with the definition $\log(0) = -\infty$.
We call $u$ {\it the energy function} for the Markov system. Recall that $u$ is not upper-semicontinuous in general \cite{Wer6} (even for contractive Markov systems with an open partition), therefore the existing theory of thermodynamic formalism is useless in our situation.
   \end{Definition}

   Observe, that with the above notation each measure $\phi_m(F(\Lambda))$ has the following form
\begin{eqnarray*}
   \phi_m(F(\Lambda))(_m[e_1,...,e_n]) = \int\limits_{_m[e_1,...,e_n]} \exp{\sum\limits_{i=0}^{n-1}u\circ S^i }d\Lambda,
\end{eqnarray*}
if $\Lambda(D) = 1$ and each $K_i$ is open in $K$,
which is similar to the Sinai's starting point for a construction of a Gibbs measure \cite{S}.

 \begin{Definition}
 Let $\nu\in P(K)$ and $m\leq 0$. Set
\[\tilde\phi_m(\nu)(A\times Q):=\int\limits_{A}P^m_x(Q)d\nu(x)\]
for all $A\in\mathcal{B}(K)$ and $Q\in\mathcal{A}_m$.
It is not difficult to see that  $\tilde\phi_m(\nu)$ extends uniquely to a probability measure on the product $\sigma$-algebra $\mathcal{B}( K)\otimes\mathcal{A}_m$ with
\[\tilde\phi_m(\nu)(\Omega)=\int
P^m_x\left(\left\{\sigma\in\Sigma:(x,\sigma)\in\Omega\right\}\right)d\nu(x)\]
for all $\Omega\in\mathcal{B}( K)\otimes\mathcal{A}_m$ and
\[\int \psi d\tilde\phi_m(\nu)=\int\int \psi(x,\sigma)dP^m_x(\sigma)d\nu(x)\]
for all $\mathcal{B}(K)\otimes\mathcal{A}_m$-measurable and
$\tilde\phi_m(\nu)$-integrable functions $\psi$ on $K\times\Sigma$ (see \cite{Wer6}).
 \end{Definition}

   Analogously to Lemma \ref{shm} one readily checks that
    \begin{equation}\label{tPhiRel}
   \tilde\phi_{m-1}(\nu)(K\times Q) = \tilde\phi_m(U^*\nu) (K\times Q)
   \end{equation}
    for all $Q\in\mathcal{A}_m$, $m\leq 0$.

  \begin{Proposition}\label{gm} Let $e\in E$ and $m\leq 0$. Then the following is true.

  (i)
        \begin{equation}\label{gmf}
            E_{\phi_m(\nu'_0)}\left(1_{_1[e]}|\mathcal{F}_m\right)(\sigma) = p_e\circ w_{\sigma_0}\circ ... \circ w_{\sigma_m}(x_{i(\sigma_m)})
        \end{equation}
        for $\phi_m(\nu'_0)$-a.e $\sigma\in\Sigma$.

  (ii) Set $p_{em}(x,\sigma) := p_e\circ w_{\sigma_0}\circ ... \circ w_{\sigma_m}(x)$ for $x\in K$ and $\sigma\in\Sigma$. Let $\nu\in P(K)$. Then
   \begin{equation}\label{mp}
            E_{\tilde\phi_{m-1}(\nu)}\left(p_{e(m-1)}|K\otimes\mathcal{F}_m\right) = p_{em}
        \end{equation}
        $\tilde\phi_{m-1}(\nu)$-a.e., where $K\otimes\mathcal{F}_m$ denotes the product $\sigma$-algebra of the trivial $\sigma$-algebra on $K$ and $\mathcal{F}_m$.
   \end{Proposition}
   {\it Proof.}
   Let $_m[e_m,...,e_0]$ be a cylinder from $\mathcal{F}_m$. Then
   \begin{eqnarray*}
       \int\limits_{_m[e_m,...,e_0]}1_{_1[e]}d\phi_m(\nu'_0) &=& \int P^m_x\left( _m[e_m,...,e_0,e]\right)d\nu'_0(x)\\
       &=& \int P^m_x\left( _m[e_m,...,e_0]\right)p_e\circ w_{e_0}\circ ... \circ w_{e_m}(x)d\nu'_0(x)\\
       &=& \int\int\limits_{_m[e_m,...,e_0]}p_e\circ w_{\sigma_0}\circ ... \circ w_{\sigma_m}(x_{i(\sigma_m)})dP^m_x(\sigma)d\nu'_0(x)\\
       &=& \int\limits_{_m[e_m,...,e_0]}p_e\circ w_{\sigma_0}\circ ... \circ w_{\sigma_m}(x_{i(\sigma_m)})d\phi_m(\nu'_0)(\sigma).
   \end{eqnarray*}
Therefore (i) is true.

For (ii), observe that, by (\ref{tPhiRel}),
 \begin{eqnarray*}
       && \int\limits_{K\times _m[e_m,...,e_0]}p_e\circ w_{\sigma_0}\circ ... \circ w_{\sigma_{m-1}}(x)d\tilde\phi_{m-1}(\nu)(x,\sigma)\\
       &=& \int\sum\limits_{e_{m-1}}P^{m-1}_{x}\left( _{m-1}[e_{m-1},...,e_0]\right)p_e\circ w_{e_0}\circ ... \circ w_{e_{m-1}}(x)d\nu(x)\\
       &=&\int U\left(  P^{m}_{.}\left( _m[e_{m},...,e_0]\right)p_e\circ w_{e_0}\circ ... \circ w_{e_{m}}\right)(x)d\nu(x)\\
       &=&\int\int\limits_{_m[e_{m},...,e_0]} p_e\circ w_{\sigma_0}\circ ... \circ w_{\sigma_{m}}(x)P^{m}_{x}(\sigma)dU^*\nu(x)\\
        &=& \int\limits_{K\times _m[e_m,...,e_0]}p_e\circ w_{\sigma_0}\circ ... \circ
        w_{\sigma_{m}}(x)d\tilde\phi_{m-1}(\nu)(x,\sigma),
   \end{eqnarray*}
as it is claimed.
  \hfill$\Box$

  \begin{Remark}\label{makc}
  Observe that equation (\ref{mp}) indicates some Martingale like behavior of functions $(p_{em})_{m\leq 0}$. In fact, by  Kolmogorov's Consistency Theorem, it is a martingale equation if $\nu$ is an invariant initial distribution for the Markov system (see Proposition \ref{omrds} (i)). This seems to indicate the truth of a more general Martingale Theorem where an adapted sequence of functions satisfies the recursive  averaging of a martingale with respect to some measures on non-decreasing $\sigma$-algebras which do not satisfy the Kolmogorov's consistency condition. Such a result has been proved in Theorem \ref{gmt}. According to it, the left hand side of (\ref{gmf}) converge  to the conditional expectation on the limiting $\sigma$-algebra almost everywhere with respect to measure $\Phi\left(\nu'_0\right)$.
   \end{Remark}

   \begin{Lemma}\label{cecl}
       Let $e\in E$. Suppose $\Phi(\nu'_0)(\Sigma)>0$ and $\Phi(\nu'_0)(\Sigma\setminus D) = 0$. Then
        \begin{equation}\label{cefr}
            E_{\Phi\left(\nu'_0\right)}\left(1_{_1[e]}|\mathcal{F}\right) = \bar p_e\circ F\ \ \
            \Phi(\nu'_0)\mbox{-a.e.},
         \end{equation}
         where $\bar p_e$ denotes the continuous extension of $p_e|_{K_{i(e)}}$ on the closure of $K_{i(e)}$ (e.g. Theorem 2, p. 190 in \cite{Bou}).
   \end{Lemma}
   {\it Proof.}
      By Proposition \ref{gm} (i) and Theorem \ref{gmt},
      \begin{eqnarray*}
          E_{\Phi\left(\nu'_0\right)}\left(1_{_1[e]}|\mathcal{F}\right)(\sigma) &=&\lim\limits_{m\to-\infty}p_e\circ w_{\sigma_0}\circ ... \circ w_{\sigma_m}(x_{i(\sigma_m)})\\
          &=& \bar p_e\left(\lim\limits_{m\to-\infty} w_{\sigma_0}\circ ... \circ w_{\sigma_m}(x_{i(\sigma_m)})\right)\\
          &=&  \bar p_e\circ F(\sigma)
      \end{eqnarray*}
  for $\Phi(\nu'_0)$-a.e. $\sigma\in D$. The claim follows.
  \hfill$\Box$

   \begin{Theorem}\label{cec}
       Let $e\in E$. Suppose $\Phi(\nu'_0)(\Sigma)>0$ and at least one of the following conditions holds true:

       (i) the Markov system is contractive, \\
       (ii) $K$ is separable and $\Phi(\nu'_0)(\Sigma\setminus D) = 0$.

       Then
        \begin{equation}\label{fr}
            E_{\Phi\left(\nu'_0\right)}\left(1_{_1[e]}|\mathcal{F}\right) = p_e\circ F\ \ \ \Phi(\nu'_0)\mbox{-a.e.}.
         \end{equation}
   \end{Theorem}
   {\it Proof.}
      By Proposition \ref{cmd}(i) and Lemma \ref{cecl},
      \begin{eqnarray*}
          E_{\Phi\left(\nu'_0\right)}\left(1_{_1[e]}|\mathcal{F}\right) =  \bar p_e\circ F\ \ \ \Phi(\nu'_0)\mbox{-a.e.}
      \end{eqnarray*}
   in both cases. Suppose there exists $e_0\in E$ such that $F(\Phi(\nu'_0))(\{\bar p_{e_0} > p_{e_0}\})>0$.
  Observe that, by the hypothesis, $F(\Phi(\nu'_0))$ is a Radon measure, in case of (i) it holds true by Lemma \ref{rpl}, in case of (ii) it is a well known fact (e.g. Theorem 7.1.7, p. 70, Vol. II in \cite{B}). Furthermore, the restriction of a Radon measure on a Borel set in a metric space is a Radon measure also. Therefore, by Lusin's Theorem (e.g. Theorem 7.1.13, p.72, Vol. II in \cite{B}), there exists a compact set $C\subset \{\bar p_{e_0} > p_{e_0}\}$ such that $F(\Phi(\nu'_0))(C)>0$ and the function $\sum_{e\in E}\bar p_e|_C$ is continuous. Then
  \[\sum\limits_{e\in E}\bar p_e(x)>\sum\limits_{e\in E} p_e(x) = 1\]
  for all $x\in C$. However, by the above,
  \begin{eqnarray*}
1&<&\frac{1}{F(\Phi(\nu'_0))(C)}\int\limits_{F^{-1}(C)}\sum\limits_{e\in E}\bar p_e(x)\circ F\ d\Phi(\nu'_0)\\
    &=&\frac{1}{F(\Phi(\nu'_0))(C)}\int\limits_{F^{-1}(C)}\sum\limits_{e\in E}E_{\Phi\left(\nu'_0\right)}\left(1_{_1[e]}|\mathcal{F}\right) d\Phi(\nu'_0)\\
    &=&1,
\end{eqnarray*}
which is a contradiction. Therefore, the claim is true.
   \hfill$\Box$

   \begin{Corollary}\label{esc}
   Suppose $\Phi(\nu'_0)(\Sigma)>0$ and it is normalized and at least one of the following conditions holds true:

       (i) the Markov system is contractive,\\
       (ii) $K$ is separable and $\Phi(\nu'_0)(\Sigma\setminus D) = 0$.

    Then  $\Phi(\nu'_0)$ is an equilibrium states for $u$ and
       \[h_{\Phi\left(\nu'_0\right)}(S) = -\int u\ d\Phi (\nu'_0).\]
   \end{Corollary}
   {\it Proof.} The claim follows from Lemma 5 in \cite{Wer6}. For completeness, we prove it here in our case.
      By Theorem \ref{cec},
      \begin{eqnarray*}
          h_{\Phi\left(\nu'_0\right)}(S) &=&-\sum\limits_{e\in E}\int E_{\Phi\left(\nu'_0\right)}\left(1_{_1[e]}|\mathcal{F}\right)\log E_{\Phi\left(\nu'_0\right)}\left(1_{_1[e]}|\mathcal{F}\right)\ d\Phi(\nu'_0)\\
          &=&-\sum\limits_{e\in E}\int\limits_{_1[e]}\log p_e\circ F\ d\Phi(\nu'_0)\\
          &=&-\int\log p_{\sigma_1}\circ F(\sigma)\ d\Phi (\nu'_0)(\sigma).
      \end{eqnarray*}
That is
\begin{equation}\label{ep}
  h_{\Phi\left(\nu'_0\right)}(S) + \int u\ d\Phi(\nu'_0)= 0.
\end{equation}
To complete the proof it remains to show that
\begin{equation}\label{mpp}
h_{\Lambda}(S) + \int u\ d\Lambda\leq 0
\end{equation}
for all $S$-invariant Borel probability measures $\Lambda$. For set
\[g_e:=E_\Lambda\left(1_{_1[e]}|\mathcal{F}\right)\]
for each $e\in E$. Then as above
\[h_\Lambda(S) = -\sum\limits_{e\in E}\int\limits_{_1[e]}\log g_e\ d\Lambda.\]
If $\Lambda(\{u=-\infty\})>0$, then $h_{\Lambda}(S) + \int u\ d\Lambda=-\infty<0$. Otherwise, observer that
\begin{eqnarray*}
   h_{\Lambda}(S) + \int u d\Lambda
   =\sum\limits_{e\in E}\int\limits_{_1[e]}\log\frac{p_e\circ F}{g_e}\ d\Lambda
   \leq\sum\limits_{e\in E}\int\limits_{_1[e]}\left(\frac{p_e\circ F}{g_e} - 1\right)\ d\Lambda.
\end{eqnarray*}
It is not difficult to check that $1_{_1[e]}(p_e\circ F/g_e - 1)\in \mathcal{L}^1(\Lambda)$ for all $e\in E$ (e.g. Lemma 2 in \cite{Wer6}). Therefore, by the pull-out property of the conditional expectation,
\begin{eqnarray*}
    &&\sum\limits_{e\in E}\int\limits_{_1[e]}\left(\frac{p_e\circ F}{g_e} - 1\right)\ d\Lambda\\
    &=&\sum\limits_{e\in E}\int g_e\left(\frac{p_e\circ F}{g_e} - 1\right)\ d\Lambda\\
    &=&\sum\limits_{e\in E}\int\left(p_e\circ F - g_e\right)\ d\Lambda\\
    &=&1-1\\
    &=&0.
\end{eqnarray*}
This implies (\ref{mpp}), as desired.
   \hfill$\Box$

      \begin{Corollary}\label{sibpm}
     Suppose $\Phi(\nu'_0)(\Sigma)>0$ and at least one of the following conditions holds true:

       (i) the Markov system is contractive, \\
       (ii) $K$ is separable and $\Phi(\nu'_0)(\Sigma\setminus D) = 0$.

        Then
          \begin{equation}\label{fr}
                    \Phi(\nu'_0)|_{\mathcal{A}_0} = \phi_0\left(F\left(\Phi(\nu'_0)\right)\right).
           \end{equation}
   \end{Corollary}
   {\it Proof.}
     The claim is a straightforward consequence of Theorem \ref{cec}. Observe that
      $E_{\Phi\left(\nu'_0\right)}\left(1_{_1[e]}|\mathcal{F}\right)\circ S^{-n}$ is $\mathcal{F}$-measurable, for all $n\in\mathbb{N}$, since $S(\mathcal{F})\subset\mathcal{F}$. Therefore, by the shift invariance of $\Phi(\nu'_0)$ and the pull-out property of the conditional expectation,
      \begin{eqnarray*}
           &&\phi_0\left(F\left(\Phi(\nu'_0)\right)\right) \left( _0[e_1,...,e_n]\right)\\
           &=&\int P^1_{F(\sigma)}\left( _1[e_1,...,e_n]\right)\ d\Phi(\nu'_0)(\sigma)\\
         &=&\int  E_{\Phi\left(\nu_0\right)}\left(1_{_1[e_1]}|\mathcal{F}\right)E_{\Phi\left(\nu_0\right)}\left(1_{_1[e_2]}|\mathcal{F}\right)\circ S...E_{\Phi\left(\nu_0\right)}\left(1_{_1[e_n]}|\mathcal{F}\right)\circ S^{n-1}\ d\Phi(\nu'_0)\\
         &=&\int  E_{\Phi\left(\nu_0\right)}\left(1_{_1[e_1]}|\mathcal{F}\right)\circ S^{-n+1}E_{\Phi\left(\nu_0\right)}\left(1_{_1[e_2]}|\mathcal{F}\right)\circ S^{-n+2}...E_{\Phi\left(\nu_0\right)}\left(1_{_1[e_n]}|\mathcal{F}\right)\ d\Phi(\nu'_0)\\
         &=&\int  E_{\Phi\left(\nu_0\right)}\left(1_{_1[e_1]}|\mathcal{F}\right)\circ S^{-n+1}E_{\Phi\left(\nu_0\right)}\left(1_{_1[e_2]}|\mathcal{F}\right)\circ S^{-n+2}...1_{_1[e_n]}\ d\Phi(\nu'_0)\\
         &.&\\
         &.&\\
         &.&\\
         &=&\Phi(\nu'_0)\left( _1[e_1,...,e_n]\right)\\
         &=&\Phi(\nu'_0)\left( _0[e_1,...,e_n]\right)
      \end{eqnarray*}
      for all cylinders $_0[e_1,...,e_n]\in\mathcal{A}_0$. The claim follows.
   \hfill$\Box$

   \begin{Proposition}\label{eoim}
   Suppose $\Phi(\nu'_0)(\Sigma)>0$ and $\Phi(\nu'_0)(\Sigma\setminus D) = 0$. Then
        \begin{equation*}
            U^{*}F\left(\Phi(\nu'_0)\right) = F\left(\Phi(\nu'_0)\right).
         \end{equation*}
   \end{Proposition}
   {\it Proof.}
      Let $f$ be a real-valued, Borel-measurable and bounded function on $K$. Let $\bar w_e$ denote the continuous extention of $w_e|_{K_{i(e)}}$ on the closure of $K_{i(e)}$ for all $e\in E$. Then, by  Theorem \ref{cec} and the shift invariance of $\Phi(\nu'_0)$,
      \begin{eqnarray*}
          U^{*}F\left(\Phi(\nu'_0)\right)(f) &=& \int\sum\limits_{e\in E} p_e f\circ w_e\ dF\left(\Phi(\nu'_0)\right)\\
          &=& \int\sum\limits_{e\in E} p_e f\circ\bar w_e\ dF\left(\Phi(\nu'_0)\right)\\
          &=& \int\sum\limits_{e\in E} p_e\circ Ff\circ\bar w_e\circ F\ d\Phi(\nu'_0)\\
          &\leq& \int\sum\limits_{e\in E}\bar p_e\circ Ff\circ\bar w_e\circ F\ d\Phi(\nu'_0)\\
          &=& \sum\limits_{e\in E}\int 1_{_1[e]}f\circ\bar w_e\circ F\ d\Phi(\nu'_0)\\
          &=& \sum\limits_{e\in E}\int\limits_{_1[e]}f\circ F\circ S\ d\Phi(\nu'_0)\\
      \end{eqnarray*}
      \begin{eqnarray*}
           &=& \int f\ dF\left(\Phi(\nu'_0)\right).
      \end{eqnarray*}
  Hence, \[U^{*}F\left(\Phi(\nu'_0)\right)(f)\leq
  F\left(\Phi(\nu'_0)\right)(f).\] Since $f$ was arbitrary and both measures have the same
  norm, they must be equal.
   \hfill$\Box$

We conclude the paper with a list of questions which it opens.

1) Are Borel measures $\Phi$ and $\Phi^*$ equal?\\
2) Is $\Phi(\nu'_0)(\Sigma)>0$ for some $\nu'_0$ for every contractive Markov system with the Feller property?\\
3) Can every equilibrium state for $u$ be obtained as a normalized $\Phi(\nu'_0)$ (observe that $\Lambda(D) = 1$ if $\Lambda$ is an equilibrium state for $u$)?

\subsection*{Acknowledgements}
I would like to thank Greg Rempala for his support, without it this work would not have been possible.

\end{document}



\title{ Erratum: Dynamically defined measures and equilibrium states}
\author{Ivan Werner}
\email  { ivan\_werner@mail.ru}
\date{20 June 2012}
\maketitle

The purpose of this note is to correct some errors which the author found in the proofs of Theorem 2 and Lemma 9 in \cite{Wer11}.  The error in the proof of Theorem 2 seems to be fatal. In any case,  the theorem is not that simple consequence of the construction of $\Phi$. However,  Lemma 9 and all the results which follow from it still can be proved under apparently stronger assumptions. These assumptions are still strictly weaker than the usual assumptions for constructions of Gibbs measures, such as summability of variation of the energy function (see Example 2 in \cite{Wer11} and Example 3 in  \cite{Wer6}). In particular, the assertion of the paper that the proposed  measure-theoretic technique allows to construct  equilibrium states in the thermodynamic sense, which minimise the free energy of the dynamical system, for some energy functions which are not upper semicontinuous is still correct.

Lemma 9 and Theorem 3 in \cite{Wer11} can be replaced with the following lemma.

   \begin{Lemma}\label{cecl}
       Let $e\in E$. Suppose $\mathcal{M}$ is contractive with $P^0_x\ll P^0_y$ for all $x,y\in K_i$ and $1\leq i\leq N$, and there exists $\mu\in P(K)$ such that $U^*\mu = \mu$. Suppose  each $K_i$ is open. Let $S:=\{i|\ \mu(K_i)>0\}$ and $\nu'_0:=1/|S|\sum_{i\in S}\delta_{x_i}$. Then
        \begin{equation}\label{cefr}
            E_{\Phi\left(\nu'_0\right)}\left(1_{_1[e]}|\mathcal{F}\right) =  p_e\circ F\ \ \
            \Phi(\nu'_0)\mbox{-a.e.}.
         \end{equation}
   \end{Lemma}
   {\it Proof.}
By the hypothesis and Proposition 4 (ii) in \cite{Wer11}, $\Phi(\nu'_0)(\Sigma)>0$.    Let $A\in\mathcal{A}_0$ be such that $\phi_0(\mu)(A)=0$. Then
$P^0_x(A) = 0$ for $\mu$-a.e. $x\in K_i$ with $i\in S$. Hence, by the hypothesis, $P^0_{x_i}(A) = 0$ for  all $i\in S$. Therefore,
 $\phi_0(\nu'_0)(A)=0$. Thus $\phi_0(\nu'_0)\ll\phi_0(\mu)$.  Hence, by Lemma 2 (ii) in \cite{Wer3}, $\Phi(\nu'_0)\ll\Phi(\mu)$. 
It is a well know fact that for any positive shift-invariant Borel measures $\Lambda$ and $M$,  absolute continuity relation $\Lambda\ll M$ implies that $E_{M}(1_{_1[e]}|\mathcal{F})=E_{\Lambda}(1_{_1[e]}|\mathcal{F})$ $\Lambda$-a.e. (as  the shift-invariance of the measures implies that  $d\Lambda/dM \circ S =  d\Lambda/dM$ $M$-a.e., where  $d\Lambda/dM$ is the Radon-Nikodym derivative, which in turn implies that $E_{M}(d\Lambda/dM|\mathcal{F}) \circ S = E_{M}(d\Lambda/dM|\mathcal{F})$ $M$-a.e., and this implies the fact). Therefore, 
\[E_{\Phi\left(\nu'_0\right)}\left(1_{_1[e]}|\mathcal{F}\right) = E_{\Phi\left(\mu\right)}\left(1_{_1[e]}|\mathcal{F}\right)\ \ \  \Phi(\nu'_0)\mbox{-a.e.}.\] 
Furthermore, by Lemma 6 in \cite{Wer6} or Theorem 1 (iii) in \cite{Wer12},
\[E_{\Phi\left(\mu\right)}\left(1_{_1[e]}|\mathcal{F}\right) = p_e\circ F\ \ \  \Phi(\mu)\mbox{-a.e.}\] 
(the requirement of Lemma 6 in \cite{Wer6} that $\sum_{i=1}^N\int_{K_i}d(x, x_i) d\mu(x)<\infty$ was not necessary, as it is true for every invariant measure $\mu$, see  Theorem 1 (ii) in \cite{Wer12}).
This implies the assertion.
  \hfill$\Box$

Finally, Corollary 1 in \cite{Wer11} can be replaced with the next theorem.

   \begin{Theorem}\label{cec}
       Let $e\in E$. Suppose $\mathcal{M}$ is contractive with $P^0_x\ll P^0_y$ for all $x,y\in K_i$ and  $1\leq i\leq N$, and there exists $\mu\in P(K)$ such that $U^*\mu = \mu$. Suppose  each $K_i$ is open. Let $S:=\{i|\ \mu(K_i)>0\}$ and $\nu'_0:=1/|S|\sum_{i\in S}\delta_{x_i}$. 
        Then  the normalized $\Phi(\nu'_0)$ is an equilibrium states for $u$ given by
 \begin{equation*}
    u(\sigma):=\left\{\begin{array}{cc}
    \log  p_{\sigma_1}\circ F(\sigma)&  \mbox{if }\sigma\in D\\
    -\infty& \mbox{ otherwise, }
     \end{array}\right.
     \end{equation*}
     with the definition $\log(0) := -\infty$,
and
       \[h_{\Phi\left(\nu'_0\right)}(S) = -\int u\ d\Phi\left(\nu'_0\right).\]
   \end{Theorem}
   {\it Proof.} The proof is the same as that of Corollary 1 in \cite{Wer11} with the only difference that instead of referring to Theorem 3 in \cite{Wer11} we refer to Lemma   \ref{cecl} above.
   \hfill$\Box$



\title{ Erratum II: Dynamically defined measures and equilibrium states}
\author{Ivan Werner}
\email  { ivan\_werner@mail.ru}
\date{June 15, 2015}
\maketitle

These corrections are required because of the errors  which the author has recently found in the proofs of Lemma 1 and Lemma 5 in \cite{Wer10} and Lemma 2 (ii) in \cite{Wer3} (see \cite{Wer13}). 

Lemma 1 in \cite{Wer10} claimed that the covering sets in the definition of the outer measure can be taken from the algebras, instead of the $\sigma$-algebras, but it was only needed for the assertion that the proposed technique is a proper generalization of  the Carath\'{e}odory outer measure, which now remains open. Everywhere in \cite{Wer10}, the restricted covers can be safely replaced with the covers from the definition (Lemma 2 gives then disjoint covers from the $\sigma$-algebras, and for the proof of Proposition 1 see the proof of Proposition 1 in \cite{Wer3} where only the covers from the definition were used) except in Lemma 5.

The latter can be replaced with the following, which shows that $\Phi$ is a construction on compact sets because, on such sets, it can be obtained with an arbitrary precision with finite covers from the algebras of finite unions of cylinder sets (it strengthens the first part of Lemma 9 in \cite{Wer15} in this case).  (We will use the notation from \cite{Wer10}.)
  
For $Q\subset\Sigma$, let $\dot\xi(Q)$ be the class of all  disjoint covers $(A_m)_{m\leq 0}\in\C(Q)$ with at most finitely many non-empty sets where each $A_m$ is a finite union of cylinder sets from $\A_m$.
 Set \[\dot\Phi(Q):=\inf\limits_{(A_m)_{m\leq 0}\in\dot\xi(Q)}\sum\limits_{m\leq 0}\phi_0\left(S^mA_m\right).\]
   \begin{Lemma} Let $Q\subset\Sigma$ be compact. Then
       \[\Phi(Q)=\dot\Phi(Q).\]
  \end{Lemma} 
    {\it Proof.} The proof is an adaptation of the proofs of Lemma 1, Lemma 2 and Lemma 5 in \cite{Wer10}.

    1. Step.  Let $\tilde\xi(Q)$ be the set of all $(A_m)_{m\leq 0}\in\C(Q)$ such that each $A_m$ is an at most countable union of cylinder sets from $\A_m $. Set
   \[\tilde\Phi(Q):=\inf\limits_{(A_m)_{m\leq 0}\in\tilde\xi(Q)}\sum\limits_{m\leq 0}\phi_0\left(S^mA_m\right).\]
 Then, obviously, $\tilde\Phi(Q)\geq \Phi(Q)$.
Let $\left((A^n_m)_{m\leq 0}\right)_{n\in\mathbb{N}}\subset\mathcal{C}(Q)$ such that
    $\sum\limits_{m\leq 0}\phi_m(A^n_m)\downarrow \Phi(Q)\mbox{ as }n\to\infty$.
Then, by  the  Carath\'{e}odory construction, for every $n\in\mathbb{N}$ and $m\leq 0$ there exists an at most countable union of cylinder sets $C^n_m\in\A_m$ such that
$ A^n_m\subset C^n_m$ and
\begin{equation*}
    \phi_m(C^n_m\setminus A^n_m)<\frac{2^m}{n}.
\end{equation*}
Therefore,
\begin{equation*}
\tilde\Phi(Q)\leq\sum\limits_{m\leq 0}\phi_m\left(C^n_m\right) \leq\frac{2}{n} + \sum\limits_{m\leq 0}\phi_m\left(A^n_m\right)~~\mbox{ for all } n\in\N.
\end{equation*}
 Hence, $\tilde\Phi(Q)\leq\Phi(Q)$. Thus  \[\Phi(Q)=\tilde\Phi(Q).\]

2. Step. Let $\xi'(Q)$ be the set of all $(A_m)_{m\leq 0}\in\tilde\xi(Q)$ such at most finitely many $A_m$'s are not empty and each $A_m$ is a finite union of cylinder sets. Set
  \[\Phi'(Q):=\inf\limits_{(A_m)_{m\leq 0}\in\xi'(Q)}\sum\limits_{m\leq 0}\phi_0\left(S^mA_m\right).\]
 Then, clearly, $\Phi'(Q)\geq \tilde\Phi(Q)$. By the compactness of $Q$, for every $(A_m)_{m\leq 0}\in\tilde\xi(Q)$ there exists $(A'_m)_{m\leq 0}\in\xi'(Q)$ such that $A'_m\subset A_m$ for all $m\leq 0$, and therefore,
\[\sum\limits_{m\leq 0}\phi_0\left(S^mA_m\right)\geq\sum\limits_{m\leq 0}\phi_0\left(S^mA'_m\right)\geq\Phi'(Q).\]
Hence $\tilde\Phi(Q)\geq\Phi'(Q)$. Thus \[\tilde\Phi(Q)=\Phi'(Q).\]

3. Step.  Obviously, $\Phi'(Q)\leq \dot\Phi(Q)$. Now, let $(A_m)_{m\leq 0}\in\xi'(Q)$, and set \[B_m:=A_m\setminus(A_{m+1}\cup...\cup A_0)\]
for all $m\leq 0$. Then, since finite unions of cylinder sets from $\A_m$ form an algebra in $\A_m$,
  $(B_m)_{m\leq 0}\in\dot\xi(Q)$, and $B_m\subset A_m$ for all $m\leq 0$. Hence
\begin{equation*}
    \sum\limits_{m\leq 0}\phi_m(A_m)\geq \sum\limits_{m\leq 0}\phi_m(B_m)\geq \dot\Phi(Q).
\end{equation*}
Therefore, $\Phi'(Q) \geq \dot\Phi(Q)$, and hence, $\Phi'(Q) = \dot\Phi(Q)$. 
\hfill$\Box$

 Lemma 2 (ii) in \cite{Wer3}  was restated in \cite{Wer10} as Lemma 3. The latter was used in  \cite{Wer10} and  \cite{Wer12} for two purposes. 

The first was to proof indirectly that the dynamically defined measure (DDM) is not zero (Proposition 3 (ii) and Proposition 4 (ii) in \cite{Wer10}). In that respect, the invalidity of Lemma 3 (which is clear by Corollary 1 in \cite{Wer15}) has been rectified to some extent in the case of contractive Markov systems (CMS) in \cite{Wer13}, through a computation of a lower bound on the norm of the DDM under stronger assumptions (see Corollary 1 in \cite{Wer13}). 

The second was for the argument that the DDM is an equilibrium state for the CMS because it is absolutely continuous with respect to one (Lemma 1 and Theorem 1 in \cite{Wer12}), where the existence of the latter is known through a non-constructive approach. That case of  Lemma 3 in \cite{Wer10} has been salvaged  in \cite{Wer15} as follows.

   \begin{Lemma}\label{dpfl}
    Suppose $\phi'_0\ll\phi_0$  and $\phi_0\circ S^{-1}= \phi_0$. Then the corresponding DDMs
\[\Phi'\ll\Phi.\] 
  \end{Lemma} 
    {\it Proof.}
See Lemma 10 in \cite{Wer15}.
\hfill$\Box$

Since the weakening of the conditions for a CMS which give non-zero DDM's seems to be likely to become a subject of ongoing research (the same way as it happened with the conditions for the uniqueness of the invariant probability measures), it may be helpful to formulate a general theorem which specifies the hypothesis on a DDM to be an equilibrium state for a CMS, so that particular theorems involving low level conditions for a CMS which at the current state of knowledge are the weakest which satisfy the hypothesis can be obtained as corollaries from the theorem.

To that end, one can take advantage of Theorem 1 in \cite{Wer11} (Theorem 3.7 in the DCDS version) which allows to extend the definition of an equilibrium state for a Markov system to that which simplifies the verification of its properties, and which can be also used in the infinite case, as follows.

 \begin{Definition}
 Let us abbreviate $\M:=(K_{i(e)},w_e,p_e)_{e\in E}$. Set 
\begin{eqnarray*}
E(\M)&:=&\left\{\Lambda\in P_S(\Sigma)|\  \Lambda(D) = 1\mbox{ and }E_\Lambda(1_{_1[e]}|\F)=p_e\circ F\right.\\  
&&\left.\Lambda\mbox{-a.e. for all }e\in E\right\}.
\end{eqnarray*}
We call the members of $E(\M)$  {\it the equilibrium states} for the CMS.
\end{Definition}

 \begin{Theorem}\label{mt}
Suppose there exists $M\in E(\M)$ such that $\phi_0\ll M|_{\A_0}$. Then $\Phi/\Phi(\Sigma)\in E(\M)$ if $\Phi(\Sigma)>0$.  
 \end{Theorem}
 {\it Proof.}
By Lemma \ref{dpfl}, $\Phi\ll M$. Hence, $\Phi(\Sigma\setminus D)=0$ (if $\Phi=\Phi(\nu_0)$ and $\M$ is contractive, than $\Phi(\Sigma\setminus D)=0$ by Lemma 3 (ii) in \cite{Wer3} (without the absolute continuity hypothesis)). Also, since $\Phi(\Sigma)>0$, it is a well-known fact that the absolute continuity relation implies that
\[E_\Phi(1_{_1[e]}|\F)=E_M(1_{_1[e]}|\F)~~\Phi\mbox{-a.e.}\]
for all $e\in E$ (as in the proof of Lemma 1 in \cite{Wer12}). Thus $\Phi/\Phi(\Sigma)\in E(\M)$.
 \hfill$\Box$

We refer to  \cite{Wer11} for the definition of the {\it non-degeneracy} of a CMS and sufficient conditions for it (e.g. the non-degeneracy is satisfied if all $K_i$' are open, but it admits also a large class of systems with proper Borel-measurable partitions). 

\begin{Corollary}
Suppose $\M$ is contractive and non-degenerate such that all $p_e|_{K_{i(e)}}$'s are Dini-continuous and there exists $\delta>0$ such that $p_e|_{K_{i(e)}}\geq\delta$ for all $e\in E$. Let $M\in E(\M)$, $S:=\{i|\ F(M)(K_i)>0\}$ and $\nu'_0:=1/|S|\sum_{i\in S}\delta_{x_i}$. Then the following holds true.

(i) $\Phi(\nu'_0)/\Phi(\nu'_0)(\Sigma)\in E(\M)$ if $\Phi(\nu'_0)(\Sigma)>0$.\\
(ii) If all $w_e|_{K_{i(e)}}$'s are contractions, then $\Phi(\nu'_0)/\Phi(\nu'_0)(\Sigma)\in E(\M)$.
\end{Corollary}
 {\it Proof.}
 (i) By Proposition 1 \cite{Wer11}, $F(M)$ is an invariant measure of the CMS, and, by Lemma 4 (i) in  \cite{Wer11} (Lemma 3.8 (i) in the DCDS version), $\Phi(F(M))=M$. Since, by Lemma 2.3 in \cite{Wer4} or Lemma 8 in \cite{Wer14}, $P_x\ll P_y$ for all $x,y\in K_i$, $1\leq i\leq N$, the integration by $F(M)$ implies that $\phi_0(\nu'_0)\ll M|_{\A_0}$ (as in the proof of Lemma 1 in \cite{Wer12}). Hence, the assertion follows by Theorem \ref{mt}.

(ii) By Corollary 1 in \cite{Wer13}, $\Phi(\nu'_0)(\Sigma)>0$. Thus the assertion follows by (i).
\hfill$\Box$

\section*{ Conclusion}
  The original claim of \cite{Wer10} that the proposed technique provides a construction for equilibrium states for CMSs is reduced to the case when all maps of the system are contractions and the CMS is non-degenerate. In this case, the local energy function associated with the system is still only measurable (the author is not aware  of any other method capable of the construction of equilibrium states for such energy functions).

\section*{Acknowledgements}
The author is grateful to Boris M. Gurevich for the invitations to give several talks on the subject at the Ergodic Theory and Statistical Mechanics Seminar at the Lomonosov Moscow State University  during the preparation to which the errors were discovered.